\documentclass{camera}
\usepackage{amssymb}
\usepackage{amsmath}
\usepackage{mathrsfs}
\usepackage{graphicx}  
\usepackage{color}

\begin{document}

%
\title{Effects of symmetry energy and momentum dependent interaction on low-energy reaction mechanisms}

%
\author{H. Zheng$^{1}$, M. Colonna$^{1}$, V. Baran$^{2}$,  S. Burrello$^{1,3}$}

%
\organization{$^{1}$ INFN-LNS, Laboratori Nazionali del Sud, 95123 Catania, Italy,\\$^{2}$ Faculty of Physics, University of Bucharest, Magurele, Romania,\\ 
$^{3}$Physics and Astronomy Dept., University of Catania, Italy.}

\maketitle

\begin{abstract}
We study the dipole response associated with the Pygmy Dipole Resonance (PDR) and the Isovector Giant Dipole Resonance (IVGDR), in connection with specific properties of the nuclear effective interaction (symmetry energy and momentum dependence), in the neutron-rich systems $^{68}$Ni, $^{132}$Sn and $^{208}$Pb. 
We perform our investigation within a microscopic transport model based on the Landau-Vlasov kinetic equation.
We observe that 
the peak energies of PDR and IVGDR are shifted to higher values when employing momentum dependent interactions, with respect to the results obtained neglecting momentum dependence.  The calculated energies are close to the experimental values and similar to the results obtained in Hartree-Fock (HF) with Random Phase Approximation (RPA) calculations.
\end{abstract}

%
\section{Introduction}
The electric dipole (E1) response of nuclei, including also exotic systems, was intensively explored during the last decade through a variety of experimental methods and probes \cite{expdata2}.
From these data,
the experimental evidence of an additional dipole strength, located below the IVGDR and dubbed as PDR, 
has emerged for several systems. 
The position of the PDR peak changes from around 7.35 MeV for ${}^{208}$Pb \cite{pb1, pb2} to 8-10 MeV for Sn isotopes \cite{sn1, sn2} and 9.55-11 MeV in the case of ${}^{68}$Ni \cite{ni1, ni3}. 


In refs. \cite{coll2, coll21},  a microscopic transport model based on the Landau-Vlasov kinetic equation 
was used to systematically study the correlation between 
the PDR peak position and  the mass number of the nucleus considered, 
and the correlation between the Energy-Weighted Sum Rule (EWSR) attributed to the 
PDR mode and the neutron skin thickness.  The role of the isovector sector of the nuclear effective interaction
was also explored, by considering three different parametrizations of the symmetry energy, but momentum dependent
effects were neglected. 
In these calculations the peak energies, for the different exotic systems considered, 
were below the experimental values, especially in the case of the IVGDR mode. In the present work, we implement the standard Skyrme interaction, including momentum dependence, into the microscopic transport model.
Three parametrizations,  the SAMi27, SAMi31 and SAMi35 \cite{coll4} are considered, in order to explore 
also the impact of the symmetry energy on the properties of
PDR and IVGDR. 

\section{Theoretical framework}

The energy density associated with standard Skyrme interactions can be expressed in terms of the isoscalar, $\rho=\rho_n+\rho_p$,
and isovector, $\rho_{3}=\rho_n-\rho_p$,  densities and 
kinetic energy densities ($\tau=\tau_{n}+\tau_{p}, \tau_{3}=\tau_{n}-\tau_{p}$) as \cite{radutaEJPA2014}:
$$
\mathscr{E}=\frac{\hbar^2}{2 m}\tau + C_0\rho^2 + D_0\rho_{3}^2 + C_3\rho^{\alpha + 2} + D_3\rho^{\alpha}\rho_{3}^2 ~+ 
C_{eff}\rho\tau + D_{eff}\rho_{3}\tau_{3} + 
$$
\begin{equation}
+ C_{surf}(\bigtriangledown\rho)^2 + D_{surf}(\bigtriangledown\rho_3)^2,
\label{eq:rhoE}
\end{equation}
where $C_i$, $D_i$ are combinations of traditional Skyrme parameters. The terms with coefficients $C_{eff}$ and $D_{eff}$ are the momentum dependent contributions to the EoS. 

\begin{figure}
\includegraphics[scale=0.4]{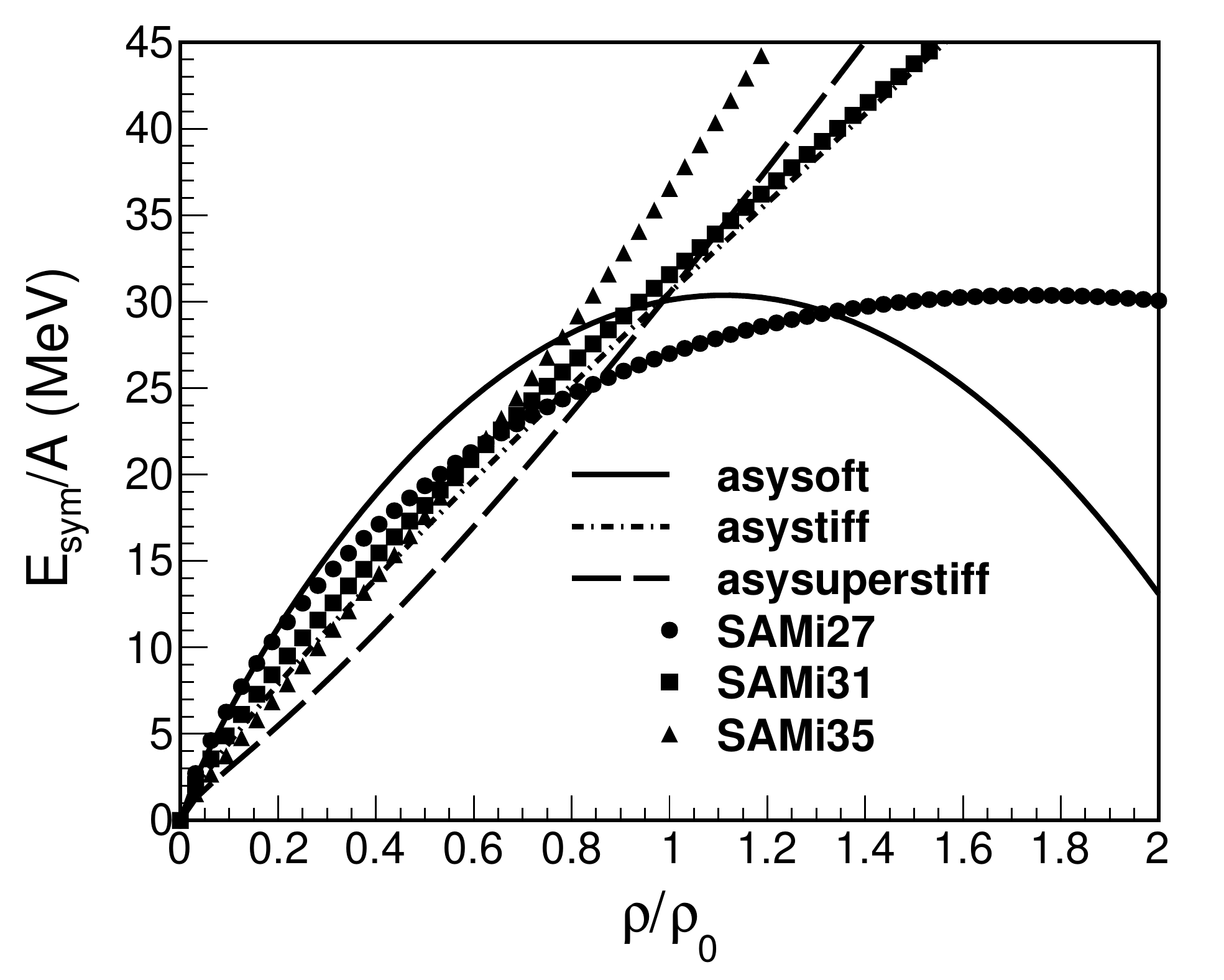}
\caption{The symmetry energy versus density for the EoS with (symbols) and without (lines) momentum 
dependent terms. 
}
\label{fig01} 
\end{figure}

Having as main ingredients the fermionic nature of the constituents
and the self-consistent mean-field, the Vlasov equation represents the semi-classical limit of Time-Dependent Hartree-Fock (TDHF) and, for small-oscillations, of the RPA equations. While the model is unable to account for effects associated with the shell structure, our self-consistent approach is suitable to describe robust quantum modes, of zero-sound type, in both nuclear matter and finite nuclei.  
The two coupled Landau-Vlasov kinetic equations for neutrons and protons:
\begin{equation}
\frac{\partial f_q}{\partial t}+\frac{\partial \epsilon_q}{\partial {\bf p}}\frac{\partial f_q}{\partial {\bf r}}-
\frac{\partial \epsilon_q}{\partial {\bf r}}\frac{\partial f_q}{\partial {\bf p}}=I_{coll}[f_n,f_p] ,
\label{vlasov}
\end{equation}
determine the time evolution of the one-body distribution functions $f_q(\vec{r},\vec{p},t)$, with $q=n,p$ \cite{baranREP2005}. 
In the equation above, $\epsilon_q$ represents the single particle energy, which can be deduced from the expression of the
energy density eq. (\ref{eq:rhoE}). 
Since we are interested in collective motion at zero temperature,  we shall switch-off the collision integral in the following.  
The parameters of the Skyrme effective interactions are fixed in order to reproduce
the saturation properties of symmetric nuclear matter, $\rho_0=0.16 fm^{-3}$, $E/A=-16$ MeV and a compressibility modulus 
$K\approx 200 - 240$ MeV \cite{baranREP2005}. 
Moreover, some specific properties of finite nuclei (radii, binding energies) can
be fitted as well \cite{coll4}.
In the case when momentum dependent terms are neglected ($C_{eff} = 0$, $D_{eff} = 0$), 
we consider, for the symmetry energy coefficient $C(\rho)$ \cite{coll2} the following expressions:
  $C(\rho) = \rho_0(482-1638 \rho)$ MeV (asysoft EoS), $C(\rho) = 36$ MeV 
(asystiff EoS) and $C(\rho) = 36 \frac{2 \rho}{(\rho + \rho_0)}$ MeV (asysuperstiff EoS).  The corresponding symmetry energy is plotted in fig. \ref{fig01} (lines). 
One can see that the symmetry energy at saturation density is almost the same for the three EoS. 
In this study, we also employed three effective interactions including momentum dependence: SAMi27, SAMi31 and SAMi35 \cite{coll4}. Since the SAMi EoS are obtained by fitting the main properties of finite nuclei, the symmetry energy is not normalized at saturation density but around $0.65\rho_0$, see fig. \ref{fig01}. 

The integration of the transport equations is based on the test-particle (or pseudo-particle) method, with a number of $1500$ t.p. per nucleon in all the cases, ensuring in this way a good spanning of the phase-space. 
From the one-body distribution functions one obtains the local densities: 
\begin{equation}
\rho_q(\vec{r},t)=\frac{2}{(2\pi\hbar)^3}\int d^3 {\bf p}f_q(\vec{r},\vec{p},t),
\end{equation}
as well as the center of mass of neutron and proton distributions, which is needed to evaluate the dipole moment:
\begin{equation}
{\bf R}_q = \frac{1}{N_q} \int \vec{r} \rho_q(\vec{r},t) d^3 {\bf r}. \label{rq}
\end{equation}

\section{Results}
It is known from quantum mechanics that the E1 response of nuclei is \cite{coll21, coll3}
\begin{equation}
S(E) = \sum_{n>0}|\langle n|\hat{D}|0\rangle|^2 \delta(E-(E_n-E_0)),
\end{equation}
where $\hat{D}$ is the dipole operator, $E_0$ is the energy of the ground state $|0\rangle$, while the energies $E_n$ correspond to the excited
states $|n\rangle$, eigenstates of the Hamiltonian of the system. The corresponding strength function in the semiclassical framework can be obtained by the Fourier transformation of the perturbation observable
\begin{equation}
S(E) = |\int_0^\infty dt~ \delta {\bf D}(t) e^{i\frac{Et}{\hbar}}|.
\end{equation}
To study the isovector dipole response, we perturbed the ground state of the nucleus by boosting the neutrons against protons in coordinate space at the beginning (but keeping the system center of mass at rest), then following the time evolution of the relative distance between the center of mass of protons and the center of mass of neutrons up to 1800 fm/c. In our transport approach: 
\begin{equation}
\delta {\bf D}(t) = \frac{NZ}{A} \left({\bf R}_p(t) - {\bf R}_n(t)\right),
\end{equation}
where $ {\bf R}_p(t)$ and $ {\bf R}_n(t)$ have been given in eq. (\ref{rq}).

\begin{figure}
\includegraphics[scale=0.3]{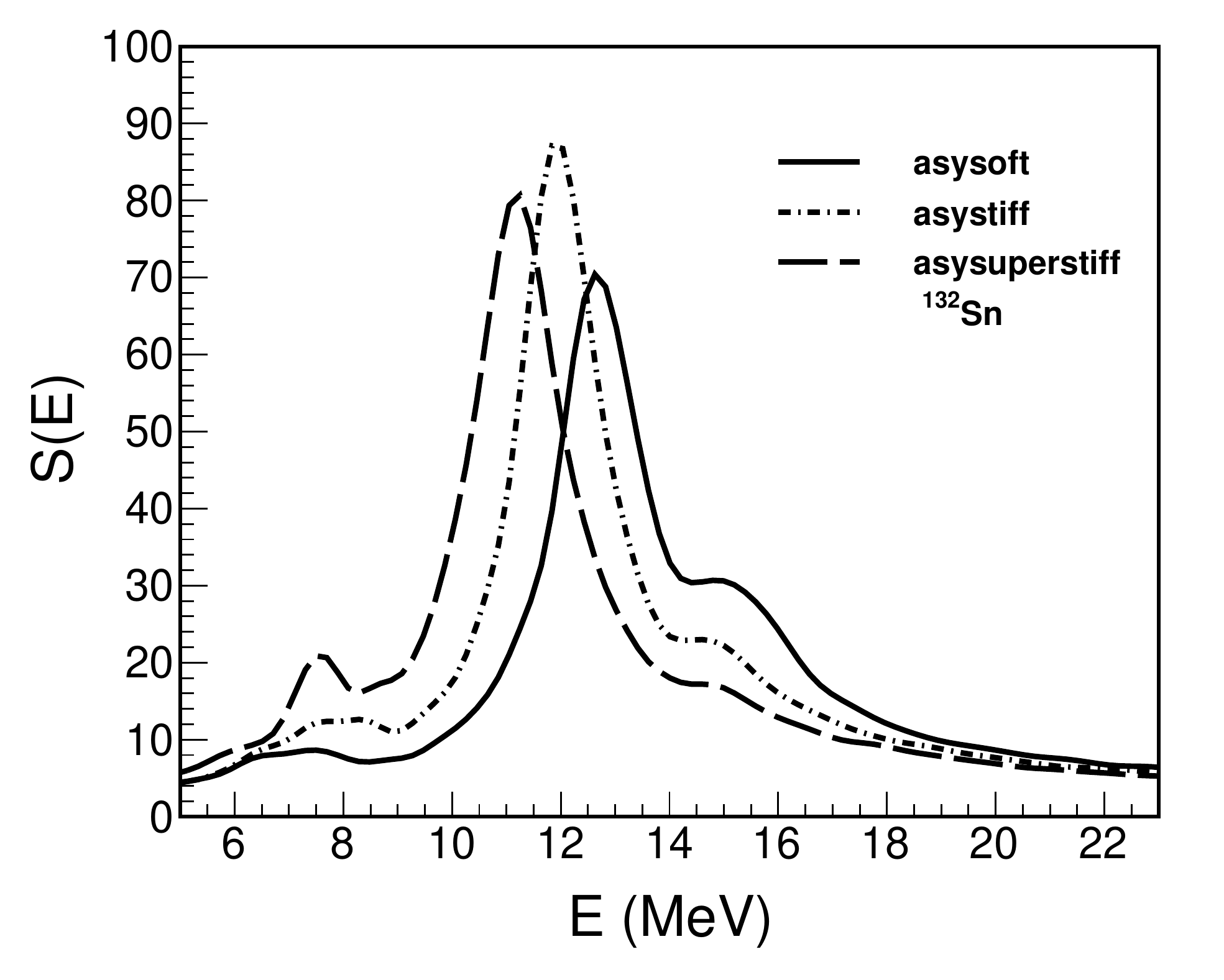}\includegraphics[scale=0.3]{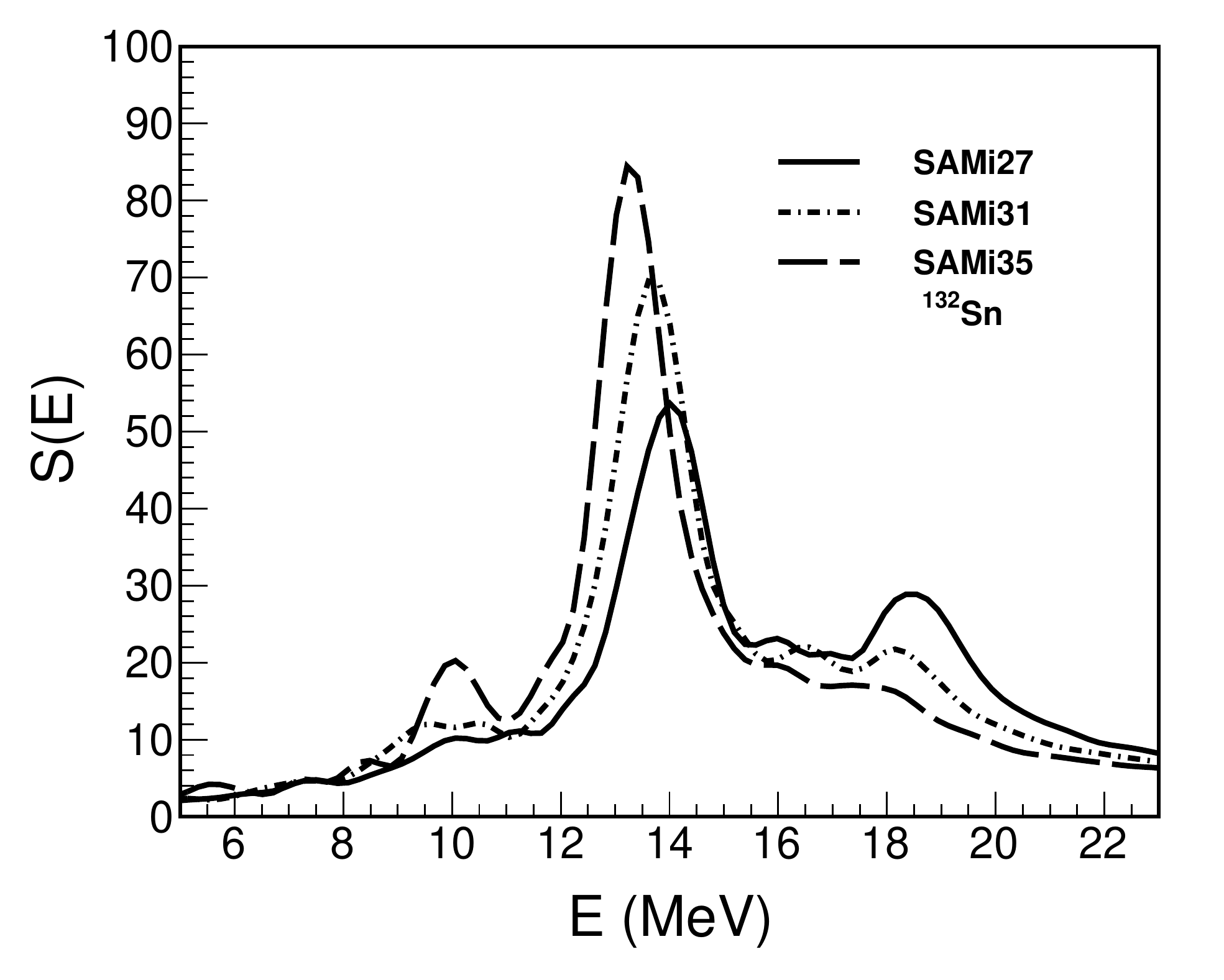}
\caption{The isovector dipole strength obtained for $~^{132}$Sn employing 
effective interactions without (left) and with (right) momentum dependent terms.} 
\label{fig02} 
\end{figure}

We performed the calculations for the three neutron-rich nuclei $^{68}$Ni, $^{132}$Sn and $^{208}$Pb, employing
the Skyrme effective interactions, with and without momentum dependent terms, introduced above. 
The results obtained for $^{132}$Sn are shown in fig. \ref{fig02}. As one can see, the peak energies of the IVGDR show a strong dependence on the EoS, when momentum dependent terms are neglected. 
Indeed, owing to surface effects, the IVGDR energies are mostly sensitive to the symmetry energy below saturation density ($\rho \approx 0.65\rho_0$), which takes different values for the three parametrizations considered
(see the lines of fig.1). For the EoS including  momentum dependence, the sensitivity of the IVGDR centroid to the symmetry energy decreases, because the three parametrizations employed cross below saturation density
(see fig.1).   
We also notice that there is a second peak above the main IVGDR peak, especially in the case of the EoS including momentum dependence, similarly to HF-RPA calculations \cite{coll4}. In the PDR region, the strength of the isovector response appears sensitive to
the symmetry energy,  both for momentum dependent and momentum independent 
interactions.  
Indeed, as pointed out in ref. \cite{coll2}, the PDR strength can be connected to the extension of 
the neutron skin, which depends on the symmetry energy parametrization, being larger
in the stiffer case.   
On the other hand, 
the energy peak of the PDR 
seems to locate at the same position, independently of the parametrization employed. 
This indicates that the PDR is mostly of isoscalar nature \cite{coll2}. Thus, to better understand the PDR features, 
it is worth investigating also the isoscalar dipole modes.  
From fig. \ref{fig02}, one can see that both the PDR and the IVGDR peak energies are shifted to higher values
when the momentum dependent terms are considered,
close to the experimental values (especially for the IVGDR) and similar to the results of HF-RPA calculations \cite{coll4}. Similar conclusions can be extracted from the study of the systems $^{68}$Ni and $^{208}$Pb. 

\section{Conclusion}
Employing several effective interactions, with and without momentum dependent terms, 
in a semi-classical transport model, we investigate the isovector dipole response
in the neutron-rich systems $^{68}$Ni, $^{132}$Sn and $^{208}$Pb. 
Our calculations indicate a clear
dependence of the peak energy of the IVGDR on the behavior of the symmetry energy below saturation density. 
On the other hand, the peak energies of the PDR do not depend on the symmetry energy, 
however the strength located in this energy region is sensitive to it.  These findings are in agreement with the results of ref. \cite{coll2}.
The peak energies of the PDR and the IVGDR are shifted to higher values 
when momentum dependent terms are included. 
The latter are close to the experimental values and similar to the results of HF-RPA calculations.



%

\end{document}